\documentclass[12pt]{article}
\begin{document}
\title{{\bf HUGE VIOLATIONS
\\ OF BEKENSTEIN'S ENTROPY BOUND}
\thanks{Alberta-Thy-06-00, quant-th/0005111}}
\author{
Don N. Page
\thanks{Internet address:
don@phys.ualberta.ca}
\\
CIAR Cosmology Program, Institute for Theoretical Physics\\
Department of Physics, University of Alberta\\
Edmonton, Alberta, Canada T6G 2J1
}
\date{(2000 May 24)}
\maketitle
\large

\begin{abstract}
\baselineskip 16 pt

	Bekenstein's conjectured entropy bound for
a system of linear size $R$ and energy $E$,
$S \leq 2 \pi E R$,
can be violated by an arbitrarily large factor,
among other ways, by a scalar field having a symmetric potential
allowing domain walls, and
by the electromagnetic field modes between
an arbitrarily large number of conducting plates.

\end{abstract}
\normalsize
\baselineskip 16 pt
\newpage

	Motivated by some considerations
of lowering objects into black holes,
Bekenstein
\cite{Bek}
conjectured that the entropy $S$
of a system confined to radius $R$
or less and energy $E$ or less
would obey the inequality
 \begin{equation}
 S \leq 2 \pi E R.
 \label{eq:1}
 \end{equation}
He found many examples obeying this inequality
\cite{Bek},
though many counterarguments and counterexamples
have also been noted
\cite{Page82,Unr}.

	For example, if the system
is one or more fields confined within a certain
region of radius $\leq R$,
with certain boundary conditions imposed
on the field(s) at the boundary of the region,
the Casimir effect can make the ground state
have negative energy and hence
the right hand side of the inequality
(\ref{eq:1}) negative,
whereas the entropy $S$ on the left hand side
can never be negative.
Alternatively, one can then construct
a mixed excited state with infinitesimally
tiny positive energy (expectation value)
but finite positive entropy,
violating Bekenstein's bound (\ref{eq:1})
\cite{Page82}.
Indeed, if
 \begin{equation}
 B \equiv {S \over 2 \pi E R},
 \label{eq:2}
 \end{equation}
so that Bekenstein's conjectured bound is
 \begin{equation}
 B < 1,
 \label{eq:3}
 \end{equation}
this example allows $B$
to be arbitrarily large.

	This particular counterexample
may be eliminated by redefining the energy $E$
to be the excess of the expectation value
of the Hamiltonian over that of the ground state
with the same boundary conditions.
But even this redefinition may not be sufficient
to prevent $B$ from becoming arbitrarily large
for interacting fields
\cite{Page82}.

	In particular, consider a scalar field
$\phi$ whose potential energy density $V(\phi)$
is symmetric in $\phi$ and has its global
minima at $\phi = \pm \phi_m \neq 0$,
and impose the condition $\phi = 0$ at the boundary
of the region under consideration.
If the region is large enough,
the energy of a classical configuration
with $\phi = 0$ at the boundary will have
a global minimum for a configuration
in which $\phi$ is near $\phi_m$ over most
of the region and then drops smoothly to 0
at the boundary.
(The region must be large enough that
the reduction in the potential energy density
from $V(0)$ to $V(\phi_m)$ integrates to more
than the increase in the ``kinetic''
or gradient energy density from the spatial
gradients of $\phi$ near the boundary.
For a spherical region, the potential energy
reduction is of the order of $[V(0)-V(\phi_m)]R^3$,
whereas the gradient energy increase is of the order
of at least $(\phi_m/R)^2 R^3$,
so the former definitely dominates if
 \begin{equation}
 R \gg \phi_m/\sqrt{V(0)-V(\phi_m)},
 \label{eq:4}
 \end{equation}
allowing a nonuniform $\phi(x^i)$ to minimize the energy.

	Classically, there is another global
energy minimum of exactly the same energy
with $\phi(x^i)$ replaced by $-\phi(x^i)$.
But quantum mechanically, there will be some
tiny tunneling rate between these two classical
configurations, so the quantum ground state
(with $E_0 = 0$ by definition)
will be a symmetric superposition of the two
classical energy minima (plus quantum fluctuations
of all the other modes).
However, there will also be an excited state
of energy $E_1$ which is the antisymmetric
superposition of the two classical energy minima
(plus other fluctuations).
For a large region, the energy excess $E_1-E_0 = E_1$
of this excited state will be exponentially tiny.
Therefore, a mixture of this state and of the
ground state can have finite nonzero entropy
(e.g., $S = \ln 2$ for a density matrix
diagonal in the energy basis and having
probabilities of 1/2 for both the ground state
and the slightly excited state, which will give
$E = {1\over 2}E_0 + {1\over 2}E_1 = {1\over 2}E_1)$,
but $2 \pi E R$ can be made arbitrarily small
as $R$ is made arbitrarily large
and $E$ becomes exponentially tiny.

	When the two classical extrema configurations
$\phi(x^i)$ and $-\phi(x^i)$ are well separated,
we can estimate that the excited state energy $E_1$
is given by some energy scale multiplied by $e^{-I}$,
where $I$ is the Euclidean action of an instanton
that tunnels between the two classical extrema
configurations $-\phi(x^i)$ and $+\phi(x^i)$.
This instanton will be a solution of the Euclidean
equations of motion of the field $\phi$ that
obeys the boundary condition $\phi = 0$ at
spatial radius $r = R$ for all Euclidean times $\tau$,
but which for $r < R$ interpolates between
$-\phi(x^i)$ and $+\phi(x^i)$ as $\tau$
goes from $-\infty$ to $+\infty$.
When the strong inequality (\ref{eq:4}) applies,
the static configuration $+\phi(x^i)$
that applies asymptotically for large positive $\tau$
is very near $\phi_m$ over almost all of the
spatial volume (except very near $r = R$), and
the Euclidean instanton is essentially a domain
wall concentrated at some Euclidean time that can
be chosen to be $\tau = 0$.

	The energy-per-area or action-per-three-volume
of the domain wall is
 \begin{equation}
 \varepsilon=\int_{-\phi_m}^{+\phi_m}\sqrt{2V(\phi)}\,d\phi,
 \label{eq:5}
 \end{equation}
and the three-volume of the Euclidean section at
$\tau = 0$ across the ball $r \leq R$ is $4\pi R^3/3$,
so the Euclidean action of the instanton is
 \begin{equation}
 I \approx {4\pi\over 3} R^3 \varepsilon.
 \label{eq:6}
 \end{equation}

	A suitable energy scale to multiply $e^{-I}$
is $R^2 \varepsilon$.  At our level of approximation,
it does not help to try to get the numerical
coefficient of the energy scale correct,
since our estimate (\ref{eq:6}) of the Euclidean action,
though being the dominant piece when it is large
in comparison with unity, has smaller corrections
that are also large in comparison with unity.
Therefore, using $\sim$ to mean that the logarithms
of the two sides are approximately equal
(up to differences that are small in comparison
with the logarithms themselves but which may actually
be large in comparison with unity,
so that the ratio of the two sides themselves
may be much different from unity), we get 
 \begin{equation}
 E_1 \sim R^2 \varepsilon e^{-I}
     \sim R^2 \varepsilon
        \exp{\left(-{4\pi\over 3}R^3\varepsilon\right)}
 \label{eq:7}
 \end{equation}
and 
 \begin{equation}
 B \equiv {S \over 2 \pi E R}
   = {\ln{2}\over\pi E_1 R}
   \sim {e^I \over I}
   \sim {\exp{\left({4\pi\over 3}R^3\varepsilon\right)}
        \over R^3 \varepsilon}
   \sim \exp{\left({4\pi\over 3}R^3\varepsilon\right)},
 \label{eq:8}
 \end{equation}
which can be made arbitrarily large by making $R$
arbitrarily large.  Indeed, $B-1$, the violation
of Bekenstein's conjectured bound (\ref{eq:3}) if
it is positive, grows large very rapidly with $R$
large enough to obey the inequality (\ref{eq:4}).

	For example, take a toy model in which
 \begin{equation}
 V(\phi) = {\lambda\over 4}(\phi^2-\phi_m^2)^2
         = {\lambda\over 4}\phi^4 - {\mu^2\over 2}\phi^2
	     + {\mu^4\over 4\lambda}
 \label{eq:9}
 \end{equation}
with, say, $\phi_m = 1$, $\lambda = 10^{-12}$, and hence
$\mu = \sqrt{\lambda}\,\phi_m = 10^{-6}$ in Planck units
($\mu = 1.221 \times 10^{13}$ GeV in conventional
high-energy physics units).  Then
 \begin{equation}
 \varepsilon = {2\over 3}\sqrt{2\lambda}\,\phi_m^3
 = \sqrt{8/9}\times 10^{-6}
  \approx 7.06\times 10^{72}{J\over m^2},
 \label{eq:10}
 \end{equation}
so for $R \gg \phi_m/\sqrt{V(0)-V(\phi_m)} =
2\lambda^{-1/2} = 2 \times 10^6 = 3.232 \times 10^{-29}\, m$,
one gets
 \begin{eqnarray}
 I & \approx & {8\sqrt{2}\,\pi\over 9}\lambda^{1/2}\phi_m^3 R^3
   \approx 9.36 \times 10^{98}
   \left({\lambda\over 10^{-12}}\right)^{1\over 2}
   \phi_m^3 \left({R\over 1\, m}\right)^3 \nonumber \\
   & \approx & 2.75 \times 10^{166}
   \left({\lambda\over 10^{-12}}\right)^{1\over 2}
   \phi_m^3 \left({R\over Mpc}\right)^3 \gg 1.
 \label{eq:11}
 \end{eqnarray}
When this Euclidean tunneling action is inserted
into Eq. (\ref{eq:8}), one gets that
Bekenstein's supposedly bounded-by-unity quantity is
 \begin{eqnarray}
 B & \sim & {e^I \over I} \sim e^I
   \sim \exp{\left({8\sqrt{2}\,\pi\over 9}
     \lambda^{1/2}\phi_m^3 R^3 \right)}
   \sim \exp{\left[ 9.36 \times 10^{98}
   \left({\lambda\over 10^{-12}}\right)^{1\over 2}
   \phi_m^3 \left({R\over 1\, m}\right)^3 \right]} \nonumber \\
   & \sim & \exp{\left[ 2.75 \times 10^{166}
   \left({\lambda\over 10^{-12}}\right)^{1\over 2}
   \phi_m^3 \left({R\over Mpc}\right)^3 \right]},
 \label{eq:12}
 \end{eqnarray}
which is utterly enormous if
$R \gg \lambda^{-1/6}\phi_m^{-1}$.
In fact, one can calculate that the violation, $B-1$,
of Bekenstein's conjectured bound is larger
than a googolplex if
 \begin{equation}
 R > \left({9\times 10^{100}\ln{10}
 \over 8\sqrt{2}\,\pi}\right)^{1/3} \lambda^{-1/6}\phi_m^{-1}
 \approx 2.91\, m
  \left({\lambda\over 10^{-12}}\right)^{-{1\over 6}}
   \phi_m^{-1}.
 \label{eq:13}
 \end{equation}

	One may now try to exclude this extreme
counterexample as well by restricting attention
to free fields and boundary conditions which lead
to unique trivial conditions (e.g., $\phi = 0$)
minimizing the energy classically.

	Then the next counterexample,
respecting the redefinition to $E_0 = 0$
and the restriction to free fields,
would simply be to consider an arbitrarily large
number $N$ of free fields.
If each field has the same lowest excited state energy
$E_1$, there are $N$ orthogonal states of this energy.
A density matrix proportional to the identity
for these $N$ states would have $S = \ln{N}$,
which for sufficiently large $N$
exceeds Bekenstein's conjectured bound
\cite{Page82}.

	The obvious next move to retain
Bekenstein's bound is also to restrict
the number of fields, say to the
approximately free fields actually observed
in nature.  But even then one can make
$B\equiv S/(2\pi ER)$ arbitrarily large
by the following density matrix diagonal
in the energy representation:
Let the state have probability $p$
to be in the excited state of energy $E_1$,
and probability $1-p$ to be in the ground state
with energy defined to be $E_0 = 0$.
Then the expectation value of the energy
is $E = p E_1$, and the entropy is
 \begin{equation}
 S = - tr\rho\ln\rho = -\sum_i p_i\ln p_i
   = -(1-p)\ln(1-p)-p\ln p,
 \label{eq:14}
 \end{equation}
so
 \begin{equation}
 B \equiv {S\over 2\pi ER}
   = {1\over 2\pi E_1 R}
      \left[\left({1\over p}-1\right)\ln{1\over 1-p}
      	+\ln{1\over p}\right].
 \label{eq:15}
 \end{equation}
As $p$ is made arbitrarily small, the last term inside
the square brackets becomes arbitrarily large and makes
$B$ arbitrarily large,
again violating Bekenstein's conjectured bound
by an arbitrary amount.

	It is not clear whether there is a natural way
further to restrict the applicability of
a conjectured bound like Bekenstein's
on the entropy-to-energy ratio to keep $B$ finite,
since even thermal states with sufficiently low
temperature always make $B$ arbitrarily large
by essentially the same reasoning,
and it seems rather unnatural
to exclude thermal states from consideration.
However, if one is determined to find a class
of states giving some bound on $B$,
one might consider those in which the density matrix
has exactly $n > 1$ nonzero eigenvalues,
all equal (and hence being $1/n$).
I.e., in a certain basis the density matrix has the
only nonzero entries being the $n$ diagonal
elements that each have the value $1/n$,
the probability that the state is each one
of the $n$ orthogonal basis pure states that
contribute nontrivially to the mixed state.
If $E_i$ for $0 \leq i \leq n-1$
is the expectation value of the energy
of each of these $n$ pure basis states,
then the expectation value of the energy
of the mixed state is
 \begin{equation}
 E = {1\over n}\sum_{i=0}^{n-1}E_i,
 \label{eq:16}
 \end{equation}
and the entropy is $S = \ln n$, so
 \begin{equation}
 B = {n\ln n \over 2\pi R \sum_{i=0}^{n-1}E_i}.
 \label{eq:17}
 \end{equation}
 
 	For a given set of fields and boundary conditions,
$B$ is maximized if the $E_i$ are chosen to be the
$n$ lowest energy eigenvalues.
Then a conjectured bound $B \leq B_{\rm max}$
for some $B_{\rm max}$
(e.g., $B_{\rm max} = 1$
for Bekenstein's conjectured bound)
is equivalent to the conjecture that
the sum of the energies of the lowest
$n$ energy eigenvalues obeys
 \begin{equation}
 \sum_{i=0}^{n-1}E_i \geq {n\ln n \over 2\pi B_{\rm max}R}.
 \label{eq:18}
 \end{equation}

	A sufficient condition for this to be true
is that, for $n > 0$,
 \begin{equation}
 E_n \geq {1 + \ln(n+1) \over 2\pi B_{\rm max}R}.
 \label{eq:19}
 \end{equation}
For a fixed system with a given number of free
quantum fields in $D$ spatial dimensions,
for sufficiently large $n$
 \begin{equation}
 E_n \sim {\rm const.} (\ln n)^{1+1/D},
 \label{eq:20}
 \end{equation}
rising faster with $n$ than the right hand side
of the inequality (\ref{eq:19}).

	Therefore, for such a fixed system,
there exists some constant $B_{\rm max}$
such that $B \leq B_{\rm max}$
for all density matrices with exactly $n > 1$
equal nonzero eigenvalues.
However, $B_{\rm max}$ will depend on the system
and need not be restricted to being 1
or less as Bekenstein conjectured.

	There are other ways to restrict
the states and/or redefine $B$ so that $B$
is bounded for each system.  For example,
instead of limiting the density matrix
to an equal mixture of $n$ states,
one could restrict the density matrix not
to have any contribution from the ground state
or states
\cite{Page82}.
Then all states that contribute
would have energies $E_i$ bounded below by
the positive energy of the first
excited state, so the denominator of
$B \equiv S/(2\pi E R)$
cannot be made arbitrarily small.
Hence this $B$ would be bounded above
for a compact set of states.
On the other hand, for sufficiently large $E$,
the entropy $S$ is bounded above by that of
thermal radiation,  which goes as
 \begin{equation}
 S \sim {\rm const.}(ER)^{D/(D+1)}
 \label{eq:21}
 \end{equation}
when $ER$ is very large, giving a $B$
that asymptotically decreases as
 \begin{equation}
 B \sim {\rm const.}(ER)^{-1/(D+1)}.
 \label{eq:22}
 \end{equation}

	Therefore, for a fixed system with the
restriction to states not overlapping
with the ground state(s),
$B$ has an absolute maximum and is bounded
above by that maximum.
For a number of free fields of order unity,
and for rather simple boundary conditions,
one would expect the maximum value of $B$
thus restricted to be generically of the order of unity.

	However, if we temporarily return to the
interacting scalar field with the potential (\ref{eq:9}),
so the first excited state has $2\pi E_1 R \sim e^{-I}/I$
with the Euclidean tunneling instanton action $I$
being given by Eq. (\ref{eq:11}),
then if the second excited state has $2\pi E_2 R \sim 1$,
we can take the mixed state with probability $1-e^{-I}$
for the energy eigenstate with energy $E_1$
and probability $e^{-I}$ for the energy eigenstate
with energy $E_2$ to get
 \begin{equation}
 B \sim I \sim 10^{99}
   \left({\lambda\over 10^{-12}}\right)^{1\over 2}
   \phi_m^3 \left({R\over 1\, m}\right)^3,
 \label{eq:23}
 \end{equation}
which can be quite large, though not exponentially large.
E.g., this $B$, for a mixed state that excludes
the ground state, can be of the order of
at least a googol when the radius $R$ of the region
obeys the inequality (\ref{eq:13}) that would
give a $B$ of the order of a googolplex
for the mixed state there that was an equal
mixture of the ground state and the
first excited state.

	On the other hand, if one took
a scalar field with three equal minima,
one would then get three energy levels exponentially
close together, a ground state that one would define
to have $E_0 \equiv 0$ and two excited states
with energies $E_1$ and $E_2$ both having exponentially
tiny energies.  Then by having a roughly equal mixture
of these two excited states, one would get a $B$
that is exponentially large (e.g., greater than
a googolplex) even with the restriction to exclude
the ground state.

	Therefore, let us return to the restriction
to free fields (with the number of species being
of the order of unity) and continue to use the
restriction to mixed states that exclude the ground
state(s) with $E_0 = 0$.

	The next point to be made is that even if one
just considers the free electromagnetic field
and excludes the ground state, one can readily find
boundary conditions in which Bekenstein's conjectured
entropy bound is violated by an arbitrarily large factor.
The idea is to use boundary conditions that correspond
to an arbitrarily large number of perfectly
conducting parallel plates.

	Between two nearby parallel perfectly conducting
static plates (giving the boundary condition that
the tangential component of the electric field
and the normal component of the magnetic field
both vanish at the surface of the plates),
electromagnetic field configurations
in which the electric field is perpendicular to the plates,
and in which the magnetic field is parallel to the plates,
approximately obey the massless 2+1 dimensional scalar
Klein-Gordon equation, with the scalar field being the
magnitude of the electric field perpendicular to the plate
(which is essentially uniform in the assumed short distance
between the plates, though it can vary
with the two transverse spatial
dimensions parallel to the plates and with time,
the 2+1 spacetime dimensions
of the effective massless Klein-Gordon equation).
Now the eigenfrequencies of these modes
are determined by the two-dimensional
spatial configuration of the nearby plates,
so the lowest eigenfrequencies
are typically of the order of the inverse of the
linear size of the plates.

	Therefore, if we have a large number, say $n$,
of thin spaces between nearby pairs of
plates that each have linear sizes
of the order of $R$, then we have of the order of $n$
modes of the electromagnetic field with frequencies
of the order of $1/R$.
Take the lowest $n$ eigenfrequencies,
and for each, construct a one-photon state
that has has the corresponding mode in its
first excited state and the other modes all
in their ground states.
These are $n$ orthogonal excited states,
each of which has an energy of the order of $1/R$.
Therefore, the mixed state that has an equal probability
of $1/n$ for each of the one-photon states
also has an energy (expectation value) $E$
of the order of $1/R$ but an entropy of $S=\ln n$.
Since $2\pi E R$ is of the order of unity,
$B \equiv S/(2\pi E R) \sim \ln n$,
which for an arbitrarily large number $n$ of plates
can be arbitrarily larger than unity
(though growing only logarithmically with $n$).

	For concreteness and more precision,
consider the case of a sequence of $n+1$ infinitely thin
conducting shells at radii $r_i$ for $1 \leq i \leq n+1$,
with $r_{n+1} = R$.  Let the $i$th region be that
with $r_i < r < r_{i+1}$, and ignore the region $r < r_1$
(i.e., set the electromagnetic field modes there to
be in their vacuum states).
If one takes the radial electric field to have the form
 \begin{equation}
 F_{01} =
  {f(r)\over r^2}e^{-i\omega t}Y_{\ell m}(\theta,\varphi)
    + {\rm c.c.},
 \label{eq:24}
 \end{equation}
where the total angular momentum of the mode
is $\ell \geq 1$,
then the radial mode function $f(r)$ obeys the equation
 \begin{equation}
 {d^2f\over dr^2}
 +\left[\omega^2-{\ell(\ell+1)\over r^2}\right]f = 0
 \label{eq:25}
 \end{equation}
and has the boundary condition $df/dr = 0$
at $r = r_i$ and at $r = r_{i+1}$.

	The lowest eigenfrequency is for the
three $\ell = 1$ modes ($m=-1$, $m=0$, and $m=+1$),
which gives the radial mode function
 \begin{equation}
 f = A({\cos \omega r\over \omega r}+\sin \omega r)
  + B({\sin \omega r\over \omega r}-\cos \omega r).
 \label{eq:26}
 \end{equation}
The boundary condition at 
$r = r_i$ and at $r = r_{i+1}$
then determine both $A/B$ and $\omega$
as functions of $r_i$ and $r_{i+1}$.
The eigenfrequency $\omega$ is then
the solution of the transcendental equation
\cite{Thomson93,Jackson75}
 \begin{equation}
 {\tan(\omega b-\omega a)\over \omega b-\omega a} =
 {1+\omega^2 ab\over
  1+\omega^2 ab-\omega^2 a^2-\omega^2 b^2+\omega^4 a^2 b^2},
 \label{eq:27}
 \end{equation}
where, for brevity, I have set $r_i = a$ and $r_{i+1} = b$.
If one defines $k \equiv \omega(b-a)$,
one can move all the resulting $\omega$-dependence
of this equation to the right side to get
 \begin{equation}
 \left({b-a\over b+a}\right)^2 = 
 {k^2\over 2+2k\cot k + k^2 +
 2\sqrt{(3+k\cot k)^2 - 12 + 4k^2}}.
 \label{eq:28}
 \end{equation}

	After some trial and error with pen and
paper and a pocket calculator, I have found
the following approximate inversion that is correct
up to relative errors of the order of
$[(b-a)/(b+a)]^8$ when that quantity is small
(when the two shells are relatively close together)
and which turned out to have a relative error (always negative)
everywhere smaller in magnitude than one part in 315:
 \begin{equation}
 \omega \!\approx\! \left[{a^2 b^2\over 2}
  \!+\! {(b\!-\!a)^4\over 15}\right]^{\!-{1\over 4}}
  \left[{3\!\cdot\! 5\!\cdot\! 7\!\cdot\! 11\!\cdot\! 13(a+b)^6 +
  (2^9 3^2 7\!-\! 3\!\cdot\! 5\!\cdot\! 7\!\cdot\! 11\!\cdot\! 13)(b\!-\!a)^6
  \over 3\!\cdot\! 5\!\cdot\! 7\!\cdot\! 11\!\cdot\! 13(a+b)^6 +
  (2^{10} 5^2\! -\! 3\!\cdot\! 5\!\cdot\! 7\!\cdot\! 11\!\cdot\! 13)(b\!-\!a)^6}
  \right]^{\!{11\over 16}}.
 \label{eq:29}
 \end{equation}
[Using Maple, I have found that the relative error is indeed very small
for small $(b-a)/(b+a)$ and then
goes to about -0.003165 at $(b-a)/(b+a) \approx 0.780$
before going back to about -0.000121 as $(b-a)/(b+a)$
approaches unity when the inner radius is taken
to become infinitesimally smaller than the outer radius.]
The last factor in square brackets, raised to the 11/16
power, varies from 1 to $(1.26)^{11/16} \approx 1.1722$,
so if one does not mind a relative error of up to
slightly more than $17\%$, one can just drop
that moderately complicated approximate correction factor,
though it does reduce the maximum relative error
by a factor of more than 50.

	Of course, for our purposes here, we
are really just interested in the case
$r_i \equiv a \approx r_{i+1} \equiv b$, say $r$,
which gives $\omega \approx \sqrt{2}/r$,
precisely what we would have gotten for
the three $\ell = 1$ modes of the massless
Klein-Gordon equation on a sphere of radius $r$.
In fact, for arbitrary $\ell$, the frequencies
in the limit $a=b=r$ are
 \begin{equation}
 \omega = {\sqrt{\ell(\ell+1)}\over r}.
 \label{eq:30}
 \end{equation}

	Let us now estimate how large $B$
can be for a mixed state (with no overlap with
the vacuum) for the electromagnetic field in the $n$
gaps between a set of $n+1$ spherical plates.
The maximum will be attained for a truncated
thermal state (truncated by leaving out the
ground state and renormalizing the excited state
probabilities so that they add up to unity)
at the appropriate temperature $T = 1/\beta$.
If $Z_0(\beta)$ is the usual partition function
when the zero-energy ground state is included,
the partition function for the truncated thermal
state will be $Z(\beta) = Z_0(\beta) - 1$,
the same as for a system in which the ground
state is absent but all the excited states are present.
Define
 \begin{equation}
 L(\beta) \equiv \ln Z(\beta) = \ln [Z_0(\beta) - 1].
 \label{eq:31}
 \end{equation}
Since the energy (expectation value) $E$
and entropy $S$ are given by the usual thermodynamic
relations $E = - dL/d\beta$ and $S = L + \beta E$,
we can readily see from the expression for $dB/d\beta$
that $B$ is maximized when
the inverse temperature $\beta$ is chosen so that
$L(\beta) = 0$ or $Z(\beta) = 1$ or $Z_0(\beta) = 2$.
At this $\beta$, one then gets
 \begin{equation}
 B \equiv {S\over 2\pi E R} = {\beta\over 2\pi R}.
 \label{eq:32}
 \end{equation}

	For a collection of electromagnetic field modes,
each with frequency $\omega_j$, one has
 \begin{equation}
 Z_0(\beta) = \exp{\left[-\sum_j{\ln(1-e^{-\beta\omega_j})}\right]}.
 \label{eq:33}
 \end{equation}
If $\beta$ is to be chosen to make this have the value 2
when there are an enormous number of modes with nearly
the same lowest frequency $\omega_j$ that all contribute
significantly to the sum, then each contribution
must be small in magnitude, so that one can approximate
each logarithm in the sum by $-e^{-\beta\omega_j}$ to get
 \begin{equation}
 \sum_j{e^{-\beta\omega_j}} \approx \ln 2.
 \label{eq:34}
 \end{equation}
Since each term in this sum is very small,
the terms with $\omega_j$ significantly larger than
the minimum value will contribute negligibly.
Thus we can ignore the $\ell > 1$ modes and consider
only the $\ell = 1$ modes.
Since there are $2\ell + 1 = 3$
of these modes for each pair of plates labeled by $i$,
we get
 \begin{equation}
 \sum_i{\exp{\left(-{\sqrt{2}\,\beta\over r_i}\right)}}
 \approx {\ln 2 \over 3}.
 \label{eq:35}
 \end{equation}

	The value of $\beta$ (and hence of $B$)
that this leads to depends on the radial distribution
of the plates.  For simplicity, assume that the $n+1 \gg 1$
plates have radii that are uniformly distributed between
0 and $R$.  Then we can approximate the sum by
an integral to get
 \begin{equation}
 \int_0^R {{n dr\over R}
 \exp{\left(-{\sqrt{2}\,\beta\over r}\right)}}
 \approx {\ln 2 \over 3}.
 \label{eq:36}
 \end{equation}
Because $n \gg 1$ implies that the solution
will have $\beta \gg R$, most of the integral will
come from $r \approx R$, and so one gets
 \begin{equation}
 {n R \over \sqrt{2}\,\beta}
 \exp{\left(-{\sqrt{2}\,\beta\over r}\right)}
 \approx {\ln 2 \over 3},
 \label{eq:37}
 \end{equation}
or
 \begin{equation}
 \sqrt{8}\,\pi B \, e^{\sqrt{8}\,\pi B} \approx {3n\over\ln 2}.
 \label{eq:38}
 \end{equation}

	For large $n$ the approximate solution for $B$ is
 \begin{equation}
 B \approx 
 {1\over\sqrt{8}\,\pi} \ln{\left({3n\over\ln n}\right)}.
 \label{eq:39}
 \end{equation}
This is a bit less than the na\"{\i}ve estimate given above
that $B \sim \ln n$, which ignored the overall numerical
factor and also ignored the $\ln n$ given in the denominator
of the main logarithm from the effect that the frequencies of the
modes depend on $r$.  [It might even be better to write this
denominator in the logarithm as $\ln(6n/\ln n)$
or $\ln{\{6n/\ln(6n/\ln n)\}}$ or $\ldots\,$,
but I shall stop at one iteration of the
approximate inversion of Eq. (\ref{eq:38}).]

	However, the main point remains true,
that the quantity $B \equiv S/(2\pi E R)$,
which Bekenstein conjectured was bounded above by unity,
can instead be made arbitrarily large (by making the number
$n+1$ of conducting plates arbitrarily large), even if one
restricts attention to a single free electromagnetic field,
defines the ground state to have zero energy,
and then excludes that state from being a component
of the density matrix (steps taken to exclude
many other counterexamples given above).
If Bekenstein's conjectured bound is to have
any applicability, one must find even further
restrictions to prevent counterexamples like
those given in this paper.

	Some of this work was done in Haiti
while awaiting the adoption papers
for our new 2.5-year-old daughter
Ziliana Zena Elizabeth.
This research was supported in part by
the Natural Sciences and Engineering Research
Council of Canada.



\baselineskip 4pt

\begin{thebibliography}{99}

\bibitem{Bek}
J. D. Bekenstein,
Phys.\ Rev.\ {\bf D23}, 287-298 (1981);
Gen.\ Rel.\ Grav.\ {\bf 14}, 355-359 (1982);
Phys.\ Rev.\ {\bf D26}, 950-953 (1982);
Phys.\ Rev.\ {\bf D27}, 2262-2270 (1983);
Phys.\ Rev.\ {\bf D30}, 1669-1679 (1984);
Phys.\ Rev.\ {\bf D49}, 1912-1921 (1994);
Phys.\ Rev.\ {\bf D60}, 124010 (1999);
M. Schiffer and J. D. Bekenstein,
Phys.\ Rev.\ {\bf D39}, 1109-1115 (1989);
Phys.\ Rev.\ {\bf D42}, 3598-3599 (1990);
J. D. Bekenstein and M. Schiffer,
Int.\ J.\ Mod.\ Phys.\ {\bf C1}, 355 (1990);
A. E. Mayo,
Phys.\ Rev.\ {\bf D60}, 104044 (1999).


\bibitem{Page82}
D. N. Page,
Phys.\ Rev.\ {\bf D26}, 947-949 (1982).

\bibitem{Unr}
W. G. Unruh and R. M. Wald,
Phys.\ Rev.\ {\bf D27}, 2271-2276 (1983); 
Phys.\ Rev.\ {\bf D42}, 3596-3597 (1990); 
M. J. Radzikowski and W. G. Unruh, 
Phys.\ Rev.\ {\bf D37}, 3059-3060 (1988);
M. A. Pelath and R. M. Wald,
Phys.\ Rev.\ {\bf D60}, 104009 (1999);
R. M. Wald,
Class.\ Quant.\ Grav.\ {\bf 16}, A177-A190 (1999);
``The thermodynamics of black holes,'' gr-qc/9912119.

\bibitem{Thomson93}
J. J. Thomson,
{\em Recent Researches in Electricity and Magnetism},
Clarendon Press, Oxford, 1873, pp. 373 ff.

\bibitem{Jackson75}
J. D. Jackson,
{\em Classical Electrodynamics}, 2nd edition,
John Wiley \& Sons, New York, 1975, p. 387.

\end{thebibliography}
\end{document}